\documentclass[sigconf]{acmart} 

\usepackage{graphics}
\usepackage{float}
\usepackage[table]{xcolor}
\usepackage[framemethod=default]{mdframed}
\usepackage{xparse}
\usepackage{listings}
\newenvironment{mybox}[1][Default Header]
{
  \begin{mdframed}[
          backgroundcolor=red!5,
          linecolor=red!5,
          linewidth=1pt,
          innertopmargin=10pt,
          innerbottommargin=10pt,
          innerleftmargin=10pt,
          innerrightmargin=10pt,
          skipabove=15pt,
          skipbelow=15pt,
          frametitle={#1},
          frametitlefont=\bfseries,
          frametitlebackgroundcolor=red!20,
          frametitleaboveskip=5pt,  
          frametitlebelowskip=5pt,
          frametitlealignment=\flushleft]
}
{
  \end{mdframed}
}

\AtBeginDocument{%
  }

\setcopyright{acmlicensed}
\copyrightyear{2018}
\acmYear{2018}
\acmDOI{XXXXXXX.XXXXXXX}
\acmConference[Conference acronym 'XX]{Make sure to enter the correct
  conference title from your rights confirmation email}{June 03--05,
  2018}{Woodstock, NY}
\acmISBN{978-1-4503-XXXX-X/2018/06}

\renewcommand\footnotetextcopyrightpermission[1]{} 

\acmConference[arXiv preprint]{arXiv preprint}
\acmYear{2025}  

\AtBeginDocument{\colorlet{defaultcolor}{.}}

\newif{\ifhidecomments}
\hidecommentsfalse
\ifhidecomments
    
\else
    
\fi

\begin{document}

\title[Reasoning About Reasoning]{Reasoning About Reasoning: Towards Informed and Reflective Use of LLM Reasoning in HCI}

\author{Ramaravind Kommiya Mothilal}
\email{ram.mothilal@mail.utoronto.ca}
\affiliation{%
  \institution{University of Toronto}
  \country{Canada}
}

\author{Sally Zhang}
\email{sallyz.zhang@mail.utoronto.ca}
\affiliation{%
  \institution{University of Toronto}
  \country{USA}
}

\author{Syed Ishtiaque Ahmed}
\email{ishtiaque@cs.toronto.edu}
\authornotemark[1]
\affiliation{%
  \institution{University of Toronto}
  \country{Canada}
}

\author{Shion Guha}
\email{shion.guha@utoronto.ca}
\authornote{Served as co-supervisors and contributed equally to this work.}
\affiliation{%
  \institution{University of Toronto}
  \country{Canada}
}


\renewcommand{\shortauthors}{Mothilal et al.}

\begin{abstract}
Reasoning is a distinctive human-like characteristic attributed to LLMs in HCI due to their ability to simulate various human-level tasks. However, this work argues that the reasoning behavior of LLMs in HCI is often decontextualized from the underlying mechanics and subjective decisions that condition the emergence and human interpretation of this behavior. Through a systematic survey of 258 CHI papers from 2020-2025 on LLMs, we discuss how HCI hardly perceives LLM reasoning as a product of sociotechnical orchestration and often references it as an object of application. We argue that such abstraction leads to oversimplification of reasoning methodologies from NLP/ML and results in a distortion of LLMs' empirically studied capabilities and (un)known limitations. Finally, drawing on literature from both NLP/ML and HCI, as a constructive step forward, we develop reflection prompts to support HCI practitioners engage with LLM reasoning in an informed and reflective way.
\end{abstract}



\begin{teaserfigure}
    \includegraphics[trim=90 100 10 70, clip, width=\linewidth]{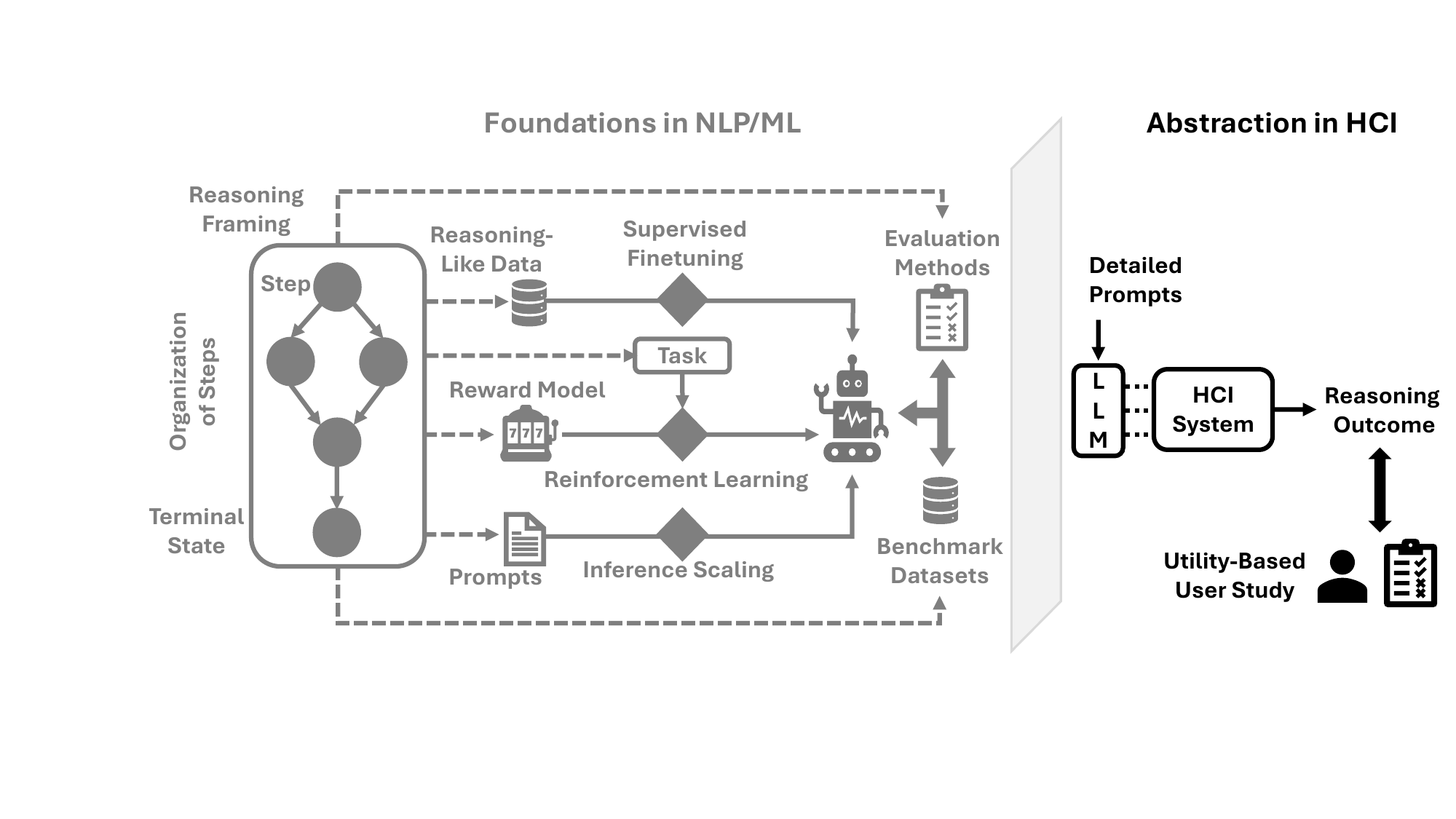}
    \caption{An overview of structural and methodological \textit{abstractions} of LLM reasoning in HCI. Within HCI, LLM reasoning is often framed as a motivator for application, executed through fine-grained prompting, and evaluated through utility-based user studies. This view abstracts away the underlying structural and methodological mechanisms that shape the emerging behaviors interpreted as reasoning by its users. Specifically, it obscures that in NLP/ML, reasoning is framed as a structured process directed toward a terminal state, executed through supervised fine-tuning or reward-based learning shaped by subjective design choices, and validated primarily through benchmark datasets.}
    \label{fig:teaser}
\end{teaserfigure}

\maketitle

\section{Introduction}
Humans \textit{reason}. Specifically, humans can identify, analyze, assess, apply and respond to reasoning. 
In HCI research, (human) users' reasoning---that is, their explanation, rationalization, or indication of why they do what they do---constitutes a central focus in understanding and enhancing their interaction with computing \cite{renom2022exploring,gao2024taxonomy,aragon2022human}.
To elicit, interpret, and analyze users' underlying reasoning behavior, a variety of theories and disciplinary inputs inform HCI research \cite{stolterman2010concept,hornbaek2017interaction,velt2017survey,clemmensen2016making}. 
In this process, what remained well-defined for many years was the boundary between users and the computing technologies in terms of their primary functions, and so, the \textit{subject} of reasoning---the users.

However, the evident surge of HCI research around Large Language Models (LLMs) in recent years has complicated the user-technology boundary. 
Prior works in HCI use LLMs across diverse domains---including communication, healthcare, education, accessibility, and design---to support a range of tasks (see \cite{pang2025understanding,inie2025co2stly,oppenlaender2025keeping} for recent surveys).
Across these domains and tasks, LLMs are employed not only as core computing technologies but also to simulate participants and users \cite{ian2025scaffold,Jin2025TeachTune}.
Further, unlike pre-LLM technologies, LLMs generate human reasoning-like responses and are continuously improved to do so \cite{tjuatja2024llms,hamalainen2023evaluating,gerosa2024can,amirizaniani2024can}.  
With this obfuscation of the boundary between the user and technology, LLMs amplify the affordance of attributing \textit{distinctive} human-like characteristics to LLM \textit{responses}.
LLMs-related research in HCI implicitly or explicitly anthropomorphizes LLMs through diverse attributions such as the ability to understand, deceive, persuade, empathize, indicate preference, motivate, and self-correct \cite{cohn2024believe,alicia2025anthro,luger2016like,Guo2025ExploringTD,Ju2025TowardAE}.
An increasingly referenced human characteristic in recent HCI studies is \textbf{reasoning}.
While both LLMs and human users are now increasingly interpreted as subjects of reasoning, the field of HCI has only paid superficial attention to the \textbf{reasoning behavior of LLMs} within the broader discourse of LLM-user interaction, which this paper aims to unpack.

To empirically understand HCI's engagement in LLM reasoning, we conduct a systematic literature review of 258 LLM-related papers published at CHI from 2020 to 2025, and ask \textbf{``how is LLM reasoning framed, executed, and evaluated in HCI?''}
We find that HCI literature that invokes the reasoning capabilities of LLMs, often by referring to or employing Chain-of-Thought (CoT) prompting technique \cite{Wei2022ChainOT}, typically focus on tasks that could be approached in parts or sequences \cite{wu2022ai,yin2025operation}, involve debate or argumentation \cite{Zhang2025BreakingBO,Zeng2025RonaldosAP,Lawley2023VALIT}, require the planning and coordination of multiple components \cite{Leusmann2025InvestigatingLC,Pan2025ACKnowledgeAC,Ma2024DynExDC}, or demand explanations and justifications \cite{zamfirescu2025beyond,Yang2023HarnessingBL,Jrke2024GPTCoachTL}.
In many studies, the reasoning-like ability of LLMs is (often implicitly) accepted with no or minimal scrutiny, and framed as the given meta-characteristic of LLMs that could motivate a specific application \cite{Deva2025KyaFP,kim2025exploring,wang2025end}.
Further, prompting the LLMs through a variety of strategies---such as including granular action-oriented instructions \cite{Lawley2024val,yin2025operation}, incorporating domain knowledge \cite{hu2024designing,li2024omniquery}, or assigning explicit personas \cite{sun2025chorus,rajcic2020mirror}---remains a gold standard approach to elicit LLM reasoning. In some cases, this is extended through the coordinated integration of multiple LLMs and external tools within a system \cite{Ge2025GenComUIEG,Earle2024DreamGardenAD,pu2025ideasynth,Lin2023JigsawSD}.
Finally, while the reasoning behavior of LLMs is often not the primary object of evaluation in HCI, it is often assessed in terms of users' subjective utility \cite{goldi2025efficient,Choi2024PrivateYS}, though this process remains obscured in some applications \cite{hintikka1996creativity,wu2024mindshift}.

Despite these increasing ways in which LLMs are becoming the subjects of reasoning, we observe that the \textbf{reasoning behavior of LLMs is only viewed \textit{partially} in HCI literature, and in particular, is decontextualized from the underlying mechanics and the subjective decisions that condition the emergence and human interpretation of this behavior.}
Therefore, we turn to the fields of Natural Language Processing (NLP) and Machine Learning (ML), where the foundations for LLMs were established, to contextualize HCI's approaches to LLM reasoning within the broader discourse on how LLMs' reasoning behavior is framed, executed, and evaluated.    
In the last few years, \textit{reasoning} has become the central focus of many studies in NLP/ML, as reflected by the growth in the number of survey works from various angles: natural language reasoning \cite{Yu2023NaturalLR}, prompting for reasoning \cite{Qiao2022ReasoningWL}, situating reasoning in agentic workflows \cite{Ke2025ASO}, and others \cite{Besta2025ReasoningLM,Luo2023TowardsLA,Xu2025TowardsLR,Sun2023ASO,huang2022towards}.
These prior works cover an overwhelming range of topics from the definitions and conceptualizations of reasoning, methods for training to reason, evaluation benchmarks for reasoning and planning, and relations to adjacent disciplines, etc.

We review this broad literature to situate HCI's engagement with LLM reasoning within the broader discourse in NLP/ML and discuss its structural and methodological abstractions from the foundational elements.    
Our analysis shows that HCI hardly perceives LLM reasoning as a product of sociotechnical orchestration, and loosely and briefly references it as an object of application.  
We argue that such abstraction leads to oversimplification of reasoning methodologies from NLP/ML and results in a distortion of LLMs' empirically studied capabilities and (un)known limitations.
Figure \ref{fig:teaser} visualizes this abstraction.
Finally, drawing on our reviews of NLP/ML and HCI literature, as a constructive step forward, we develop a conceptual tool with reflection prompts to support HCI researchers and practitioners in approaching LLM reasoning in an informed and reflective way.

In summary, \textbf{this work grounds HCI research in the conceptual trajectory of LLM reasoning in NLP/ML to ensure informed human-computer interaction systems design and evaluation involving LLMs.} We organize our contributions as follows:
\begin{itemize}
    \item Contribute a systematic literature review of 258 LLM-related full papers published in the CHI proceedings in the last six years (2020-2025) to understand how reasoning behavior is approached in the broader discourse of LLM applications in HCI (section \ref{sec:hci}).
    \item Contextualize HCI's approaches to LLM reasoning in the fields of NLP and ML, expand on the implicit assumptions in HCI, detail what HCI practitioners need to consider when using LLMs, and discuss the opportunities for HCI research in the space of LLM reasoning (sections \ref{sec:hci-abs}).
    \item Build on the above discussion to develop a conceptual tool with reflection prompts for an informed and reflective use of LLM reasoning in HCI (section \ref{sec:checklist}).
\end{itemize}




\section{Background}
\label{sec:background}

\noindent \textbf{LLM Reasoning in NLP and ML.} 
We do not adopt or offer an ontological account of reasoning itself in this work, but rather use the notion ``LLM reasoning'' to refer to the \textit{behavior} in the LLM response that is \textit{interpreted} as reasoning by its users.
This viewpoint accommodates the heterogeneous understandings of LLM reasoning across NLP and ML. 
For example, while some prior works consider the LLM output behavior as reasoning only if the LLM had faithfully followed its stated intermediate steps \cite{lanham2023measuring,paul2024making,li2024towards}, others attribute reasoning based on how convincing or plausible the output is to target humans \cite{agarwal2024faithfulness,chiang2023can}.
It is also important to note that the expected reasoning behavior could be explicitly encoded in an LLM's output, such as in terms of human-interpretable intermediate steps, or the reasoning behavior may be attributed to the LLM if the users interpret its outputs as if they resulted from reasoning required for a task.

Further, in the context of reasoning, prior works in NLP/ML often treat LLMs as \textit{systems} since a reasoning behavior can be exhibited by a standalone LLM \cite{yasunaga2023large,wang2023guiding} or through interaction with external tools, APIs, or other LLMs \cite{hammane2024selfrewardrag,yin2024mumath}. Multiple instances (or copies) of the same LLM can also be staged together to produce reasoning \cite{feng2025one}.
In multi-entity settings, an LLM integrated within a system may be described as an \textit{agent} when it is trained to perform some actions independently of human intervention, though this reflects a functional characterization rather than an ontological claim \cite{kapoor2024ai}.
Such reasoning-exhibiting LLM systems are increasingly referred to as Reasoning Language Models (RLMs) in recent works \cite{Besta2025ReasoningLM,huang2022towards,Ke2025ASO}; these RLMs differ from the earlier instances of LLMs since they are not only trained to generate the next probable tokens but also involve an orchestration of multiple LLMs or external tools and/or trained on reasoning-like datasets to exhibit reasoning.
We refer readers to \citet{Besta2025ReasoningLM} and \citet{Ke2025ASO} for recent comprehensive reviews of LLM reasoning in NLP and ML.
In this work, we refer to this vast literature to establish the grounding necessary to situate the current scope of reasoning discussion in HCI, drawing attention to its limitations and identifying opportunities for future work.

\smallskip
\noindent \textbf{Large Language Models in HCI.}
In a 2016 CHI paper, \citet{oulasvirta2016hci} characterized HCI research as a problem-solving field concerned with the human use of computing, emphasizing primarily empirical and constructive contributions.
More recently, the emergence of LLMs has significantly transformed this orientation, as evident in the increasing number of publications centered around LLMs.  
\citet{pang2025understanding} analyze 153 LLMs-related CHI papers from 2020-2024 and find more than 400\% increase in the number of papers on LLMs from 2023 to 2024.
Importantly, more than 98\% and 60\% of these studies focus on empirical and artifact contributions, a trend reflecting how HCI is rapidly adopting LLMs as a foundation for system-building
In addition, LLMs are being increasingly used in distinct roles, unimagined before, as participants, users, and performing research tasks.
These works use LLMs across diverse domains, such as communication and writing \cite{fu2023comparing,zhang2025friction}, healthcare \cite{ramjee2025ashabot,jorke2025gptcoach,calle2024towards}, education \cite{analogyUnlock,lee2024opensesame}, accessibility \cite{li2025generative,anna2024mindtalker}, design \cite{zheng2025evalignux,Lin2023JigsawSD,tang2025llm}, and software development \cite{Jiang2022DiscoveringTS,Ma2024DynExDC,Epperson2025InteractiveDA}, to perform a range of tasks.
Some of these tasks include generating research ideas \cite{Choi2025GenParaET,pu2025ideasynth}, creative writing \cite{qin2025toward,Chakrabarty2024ArtOrArtifice}, planning and decomposition of complex operations \cite{Earle2024DreamGardenAD,zhang25_chi,Lee2025VeriPlanIF}, extracting relevant information \cite{park2025leveraging,Ding2024LeveragingPL}, responding by adopting a persona \cite{Benharrak2023WriterDefinedAP,Hedderich2024APO}, and aiding multimodal interactions \cite{Guo2024WhatSM,zheng2025customizing}, among others. 

Recent works have also highlighted several concerns related to their negative outcomes, unintended effects, and research validity \cite{Shin2024FromPT,Ma2024EvaluatingTE,Zhou2023SyntheticLU}.
For instance, \cite{Inie2025HowCI} audit the use of generative AI technologies in 282 papers submitted to CHI 2024, and raise concerns over their carbon footprint and non-transparent use.
Several other studies discuss how LLMs perpetuate racial, gender, and cultural stereotypes in interactive applications, such as chatbots, and generate problematic outputs, including misinformation, hallucinations, and hateful content (see \cite{pang2025understanding} for a recent survey).
Despite these efforts, most of these works' focus largely concentrates on downstream consequences, often without a deeper conceptual analysis of the factors motivating their use.
One such factor is the attribution of ``reasoning'' abilities to LLMs---a characteristic that strongly shapes how and why they are used, yet remains underexamined within HCI.
To our knowledge, HCI has not systematically examined this presumed reasoning capacity, despite its central role in motivating adoption and obscuring critical evaluations of LLMs' capabilities.
\textbf{We address this gap in this work by not only conducting a survey of HCI literature but also by situating HCI's engagement with LLM reasoning \textit{in relation} to its foundations in the NLP/ML literature.}

\section{Methods}
In this study, we focus on \textit{generative} LLMs based on transformer architecture due to their widespread use in HCI and other related disciplines.
Though LLMs are used in many HCI-related disciplines, such as in the context of collaborative work and AI fairness, we restrict our analysis to CHI papers for their significant impact on HCI research and their representation of diverse domains and methodological approaches, following the rationale of \citet{pang2025understanding}.

\smallskip
\noindent \textbf{Data Collection.}
We relied on the ACM Digital Library to collect full-text LLM-related papers published at CHI from 2020 to 2025. 
We used the following keywords---adopted in \citet{pang2025understanding} in their recent survey on \textit{LLM-ification} of CHI---on papers' title and abstract directly on ACM Digital Library search: ``language model'' OR ``llm'' OR ``foundation model'' OR ``GPT'' OR ``ChatGPT'' OR ``Claude'' OR ``Gemini'' OR ``Falcon''.
This filter resulted in an initial corpus of 261 LLM-related papers. We manually read through each paper to verify if they involve significant discussion on LLMs, and removed 3 papers that were false positives (they either mentioned ``LLMs'' briefly in one or two sentences or used non-generative language models such as BERT).
Our final corpus consists of 258 CHI papers on LLM.
We did not filter for ``reasoning'' since it can be conceptualized, expressed, and referred to in diverse ways (more on this below), making it difficult to identify reasoning-related papers using keywords alone without reading the full text to understand the context.
We tried adding several variants and closely related concepts to ``reasoning'' to our search, but it missed out several true positive papers. So we restricted to filtering on LLM alone.

\smallskip
\noindent \textbf{Data Analysis.}
We carried out an abductive inferential process, which is recursive and iterative by definition, to analyze our corpus \cite{peirce1934collected,fann2012peirce,timmermans2012theory}.
In particular, the first author determined the initial codes for our research question based on how LLM reasoning is imagined and interpreted in NLP/ML and in consultation with co-authors.
These codes also included two categories previously used by \citet{pang2025understanding} on LLMs to identify application domains and the roles of LLMs.
Then, a set of 10 papers was randomly selected by the first author in each of five iterations to independently apply and update the codes. Each iteration maintained a 30–70\% split between papers published up to 2024 and those from 2025, as 2025 accounted for approximately 58\% of all LLM-related papers in our corpus.
While we use relevant structural and methodological elements from NLP and ML---drawing on recent surveys \cite{Besta2025ReasoningLM,Ke2025ASO,Sun2023ASO,yu2024natural}---as foundations to understand HCI's engagement with LLM reasoning, we also allowed unique orientations and approaches from HCI literature to emerge inductively. 
For example, some codes, such as those identifying the use of multi-step prompting, were informed by NLP/ML literature, whereas others, like codes capturing the use of LLMs as motivators, emerged during this initial analysis.
After these five iterations, the first author consolidated the emergent updates and refined the coding scheme in consultation with all authors and in reference to NLP/ML literature, resulting in the final set of granular codes. 

To identify LLM reasoning–related papers, we considered not only a paper's explicit mention or engagement with the reasoning behaviors or capabilities of LLMs, but also other forms of reasoning grounded in NLP/ML literature, which itself draws inspiration from philosophical accounts of reasoning \cite{yu2024natural}.
We also contrasted the presence of reasoning discussions with tasks in which LLMs do not typically perform reasoning, as interpreted in NLP/ML, or where reasoning can be reduced to simple pattern matching from the training distribution, such as summarization, factual Q\&A, or sentiment classification\footnote{While this is a debated topic, NLP/ML research typically focuses on specialized tasks such as planning and creativity to evaluate LLMs' reasoning abilities, as many other tasks can be performed through simple statistical pattern matching.}.
Figure \ref{fig:year} shows the number of reasoning-related LLM papers at CHI over the years.
Finally, the first and second authors reviewed the remaining papers in two equally sized rounds. After the first round, we randomly selected 30 papers and calculated inter-rater reliability using Krippendorff's alpha to assess consistency and resolve any disagreements. We then completed the codebook in the second round. The Supplementary Materials contain all codes and a detailed description.




\section{LLM Reasoning in HCI}
\label{sec:hci}

We now present our analysis of the HCI literature to examine its focus and contributions on LLM reasoning. Specifically, we ask how LLM reasoning is framed, executed, and evaluated in HCI.

\subsection{Framing of Reasoning}
\label{sec:hci-frame}

\begin{figure}[t]
  \includegraphics[trim=10 10 5 5, 
  clip, width=\columnwidth]{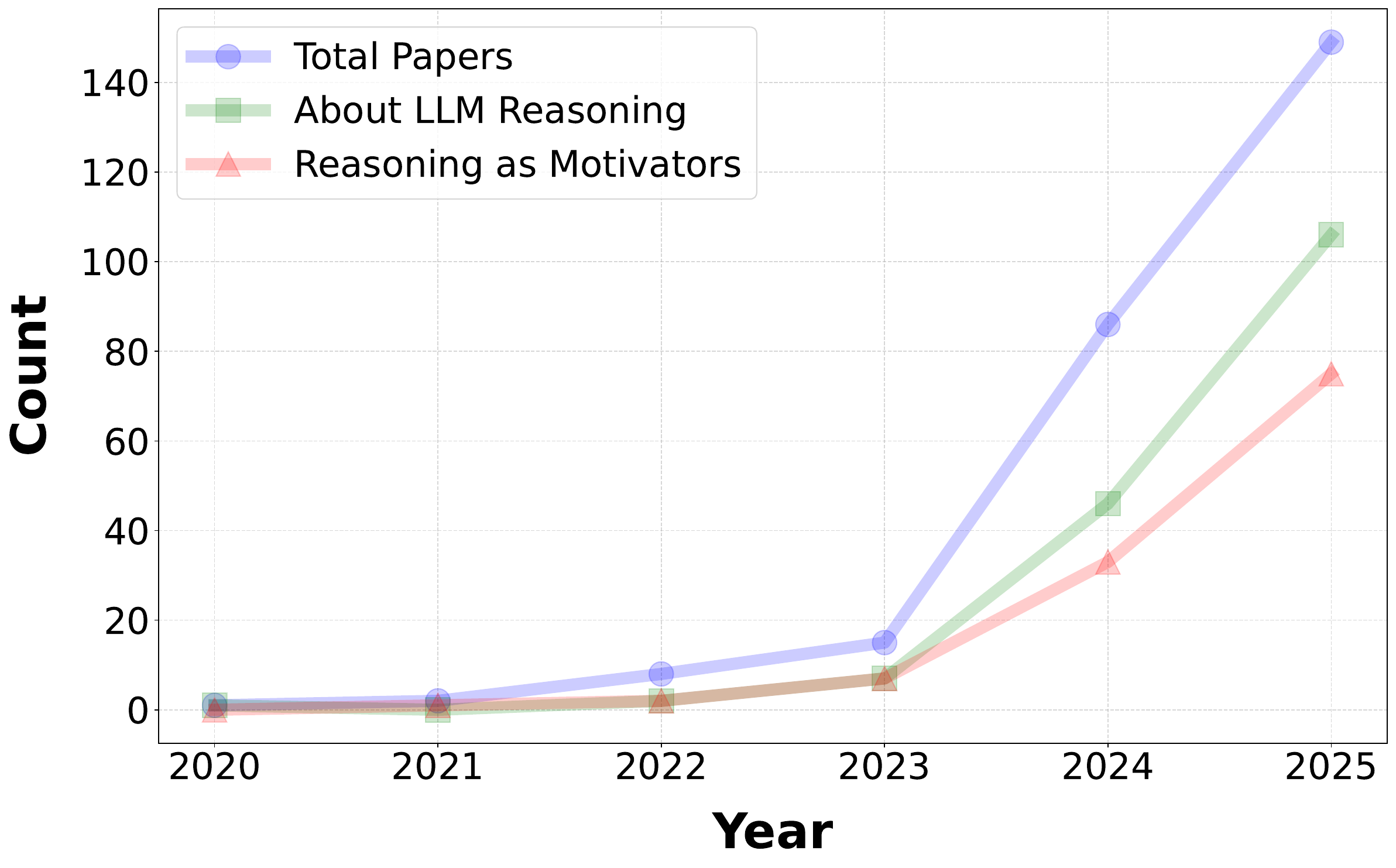}
  \caption{Total LLM-related papers at CHI over the years, including the total that engage with LLM reasoning and the total that use LLM reasoning as motivators.}
  \label{fig:year}
\end{figure}

HCI literature on LLMs often follows a tokenistic mention of LLM reasoning for various reasons, including for critiquing LLMs' unreliability or for studying users' perceptions regarding a specific task. 
We find that the reasoning capabilities and behaviors of LLMs are explicitly mentioned in 63\% (N=162) of the papers we reviewed.
In the remaining studies, 27\% allude to some form of reasoning or a closely related concept, as discussed below. 
Figure \ref{fig:year} highlights the increase in total papers discussing some aspect and form of LLM reasoning.
Across these works, we find two broad ways of framing LLM reasoning.

\subsubsection{Motivators for Application.}
\label{sec:hci-frame-motiv}
(117, 45\%) papers in our survey frame LLMs' reasoning capabilities, by referring to prior observations and results, as \textit{motivators} necessary to apply LLMs for their specific tasks. 
Many of these invocations are generic and exist across a range of tasks such as personal coaching and support \cite{goldi2025efficient,li2024omniquery}, emotional support and well-being \cite{zheng2025customizing,tang2024emoeden}, creative writing \cite{fan2024contextcam,shaer2024ai}, and ideation \cite{liu2024ai,choi2025proxona}.
For example, \citet{wang2024human} augment LLM reasoning with humans' for diverse annotation tasks as \textit{``The first one is to leverage the LLM’s capability to generate fluent and supportive explanations for its predictions that help rationalize its reasoning process.''}
To aid data wrangling in programming, \citet{chen2025dango} motivate LLM application as \textit{``Dango leverages the intrinsic reasoning capability of LLMs to proactively detect potential ambiguity in user input and ask clarification question.''}

On the other hand, some studies motivate through specific reasoning behaviors or capabilities, such as commonsense \cite{yang2025geniewizard,noh2025biassist}, logical \cite{yin2025operation}, or other domain-specific reasoning \cite{rajashekar2024human,yang2023harnessing}. 
For example, while \citet{ju2025toward} invoke analogical reasoning to facilitate empathy in microaggression situations, \citet{wang2025exploring} rationalizes their sleep health-supporting application using LLMs' reasoning capabilities emerging from finetuning on behavior change theories. 
Similarly, \citet{shao2025unlocking} use LLMs to explain scientific concepts to students by drawing on LLMs' cross-domain analogical reasoning capabilities. 
\citet{huang2024plantography} dedicate an entire section to motivate their study based on LLM reasoning and view their task of landscape image rendering from textual descriptions through different forms of reasoning---spatial, perspective relational, and commonsense reasoning.

While LLMs' reasoning characteristics may not be inherently required for tasks such as text classification and detection, we found that several studies primarily revolving around these tasks anyway motivate their systems based on LLMs' reasoning abilities \cite{kretzer2025closing,kim2025exploring,wang2025end}.
A case in point is \citet{kretzer2025closing} who motivate LLM application for a binary classification task---to detect if a user story exists in a GUI---by emphasizing that they \textit{``exploit the text understanding and reasoning capabilities of recent LLMs.''}
Other tasks where LLM generally do not perform reasoning (as interpreted in NLP/ML)---but still used as motivators in many studies---include text summarization \cite{song2025exploreself}, question answering \cite{wang2023enabling}, semantic extraction \cite{yoo2025elmi}, or text rephrasing \cite{wang2025end}, and most of these studies perform multiple such non-reasoning-focused tasks \cite{zhang2025summact,wang2024farsight}.
\citet{wang2023enabling}, for example, motivate the descriptive text summarization of essential functional purposes of mobile screens through multi-step reasoning abilities of LLMs.
Similarly, \citet{duan2024generating} invokes GPT-4's reasoning abilities to rationalize its application to perform heuristic evaluation of user interfaces and rephrasing of generated feedback. 

\subsubsection{Sequential Progression of Responses.}
\label{sec:hci-frame-seq}
Chain-of-Thought (CoT) \cite{Wei2022ChainOT} prompting elicits reasoning-like outcomes in many cases and has been applied successfully not only in HCI but also in many other disciplines.
In our review, we found that 26\% (N=47) of the studies that apply LLMs employ some form of CoT and consequently adopt a linear chain reasoning frame.
Reasoning in these studies is thus often interpreted and framed as the sequential progression of steps or components corresponding to intermediate desired responses to the given problem.
This trend is particularly apparent in Creativity and Design domains where 43\% and 33\% adopt such linear framing using CoT. 
Several works use chained step-by-step reasoning to synthesize research or design ideas \cite{zamfirescu2025beyond,pu2025ideasynth,xu2025productive,liu2024ai}, simulate interactions of target user personas \cite{xiang2024simuser,choi2025proxona}, and navigate user interfaces and extract relevant insights \cite{zhang2025appagent,park2025leveraging,yang2025geniewizard}, among others. 
For example, \citet{xiang2024simuser} uses CoT to mimic how diverse target users of a mobile application would sequentially perceive and interact with its interface and offer five different types of usability feedback.
Similarly, \citet{hou2024c2ideas} use CoT to split the interior color design process into sequential steps on which domain knowledge and user feedback are integrated to produce creative color design.

In many prior works, we found that though the task for an LLM is broken down into multiple steps, an individual reasoning step is often dense---overloaded with many sub-tasks and abstract instructions \cite{taeb2024axnav,wu2022ai,yin2025operation,chen2024empathy,zhang2025friction}.
While this can partially be attributed to the complexity of the task, oftentimes this arises as a consequence of including levels of detail (of instructions) required for the task, which also appears to result in better desired outcomes.
Consider the work of \citet{zhong2025screenaudit}, who prompt an LLM to identify and analyze potential accessibility concerns of mobile applications based on screen reader transcripts: Here, each reasoning step packs several questions, such as \textit{``...What is the intent of each transcript entry? Is the intended meaning conveyed through the transcript? Build your analysis on top of Step 1...''}, and include abstract instructions such as ``Look at the big picture. Consider each transcript entry in relation to the elements before and after it...''
In a few other studies, a reasoning step also refers to an independent component in the path to the final response, requiring its own specific prompt instructions, for instance \cite{das2025ai,viswanathan2025interaction}.

\subsection{Execution of Reasoning}
\label{sec:hci-exec}
In HCI, the reasoning behavior of LLMs is predominantly implemented through fine-grained prompting strategies, building on domain insights. Further, in many works, LLM reasoning is also interpreted to emerge through the interactions between multiple components of an HCI system in which LLMs are embedded. 

\subsubsection{Prompting as Gold Standard.}
\label{sec:hci-exec-gold}
The most apparent observation in our survey was HCI's almost complete reliance on prompting strategies when using LLMs (N=159, 90\%). By drawing on domain knowledge and/or formative studies, prior works exert significant effort to craft highly detailed and fine-grained prompts, many times with demonstrations containing specific reasoning processes. 
However, only a very few prior works include explanations on why prompting is chosen over finetuning LLMs for reasoning; those who do mostly cite convenience and presumed high returns as reasons \cite{Zhang2025AppAgentMA,jorke2025gptcoach,wang2023enabling,Zhou2023InstructPipeGV}.
While a few other works also rationalize why finetuning worked better for some of their tasks \cite{Ko2023NaturalLD,huang2024plantography,Li2024OmniActionsPD}, prompting, or inference-scaling to be precise\footnote{The term ``inference scaling'' refers to allocating additional processing time to an LLM by augmenting the input prompt with detailed instructions, explanations, and demonstrations before response generation.}, is often considered the gold standard even in these works.

Several works follow an piecemeal action-oriented approach where the prompts include a sequential list of presumably simpler tasks of a reasoning process (for e.g., \cite{kim2025exploring,wang2025social,fan2025litlinker,Lee2025VeriPlanIF}), each of which the LLMs could complete with their general text-generation capabilities, without the need for specific domain knowledge\footnote{While LLMs' capabilities are indeed due to their large-scale training on diverse domains, what we refer to here is the absence of domain-specific knowledge in the prompts.}. 
For example, \citet{zhou2025journalaide} provides the instructions in three straightforward steps to ask gentle conversational questions to older adults in recalling their experiences for digital journaling.  On the other hand, \cite{yin2025operation} includes a long sequence of sub-tasks to identify and analyze a user's UI operations, such as informal recall and comprehension, where each sub-task is provided with definite options to choose from..
Such tasks are performed to achieve a range of goals, such as health self-management \cite{mahmood2025voice}, analyzing nutritional information \cite{szymanski2024integrating}, identifying and explaining biases in news articles \cite{noh2025biassist}, and tailored programming learning assistance \cite{ma2025dbox,jin2024teach}.
The interpretation of reasoning here emerges through the act of replicating the steps of a reasoning process provided as detailed instructions in the prompt, and these instructions and steps are often determined by formative or preliminary studies.

In many other studies, domain knowledge is explicitly encoded in the prompt, since it might not be invoked without its explicit inclusion, to align LLM reasoning with domain-specific processes.
For example, to screen neurocognitive disorders through conversations, \citet{hu2024designing} utilizes scaffolding theories and practices such as Zone of Proximal Development \cite{vygotsky1978mind} to prompt LLMs to imitate specific human skills.
Similarly, \citet{wu2024mindshift} develop their prompts to nudge users against problematic smartphone usage based on several persuasion strategies grounded in theories such as the Dual Systems Theory \cite{vygotsky1978mind}.
The required knowledge can sometimes be retrieved from relevant knowledge bases and included in the prompt \cite{wang2025social,cox2021directed,ashby2023personalized}, such as the study by \citet{li2024omniquery} whose system performs a multi-source retrieval of memorable instances from users' digital libraries, extracts atomic semantic knowledge, and integrates them into the prompt to an LLM.

Another common prompting strategy is to assign explicit \textit{personas} to LLMs and nudging them to reason through the proxy human-like characteristics of these personas.
Instead of being task-oriented (for e.g., ``$\dots$your role is to understand the cause-effect relationship$\dots$'' \cite{ding2024leveraging}), personas contain core identities \cite{sun2025chorus}, communication styles \cite{lima2025promoting}, emotional traits \cite{lo2025d,rajcic2020mirror}, values and worldviews \cite{zhou2024eternagram}, and other personality characteristics relevant for the task.
Nonetheless, the above-discussed strategies---through fine-grained step-by-step instructions, explicit inclusion of domain knowledge, and assigning personas---require significant cognitive effort and resources (for formative studies or other preliminary experiments) from the researchers' end to evoke LLM reasoning to directly meet end users' needs. On the other hand, a few prior works develop interactive systems to enable users to optimize their prompts on their own to interpret LLMs' reasoning \cite{arawjo2024chainforge,kim2024evallm,wang2025end}.
For instance, \citet{he2025prompting} developed an Excel extension providing detailed instructions to an LLM in analyzing LLM-annotated labels and reasoning for tasks without ground truth data.

\begin{figure*}
  \includegraphics[trim=10 10 5 5, 
  clip, width=0.85\textwidth]{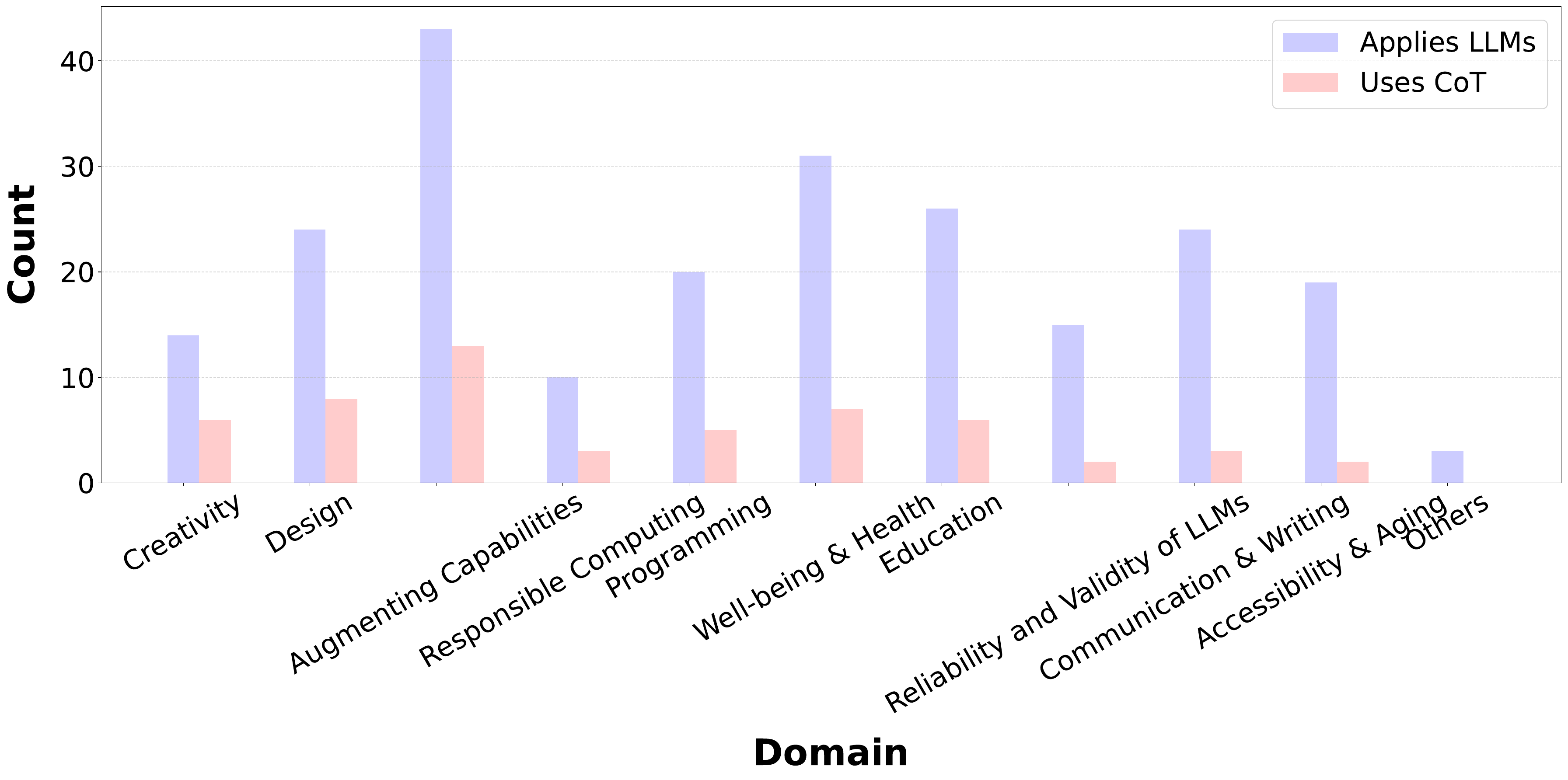}
  \caption{Total LLM-related papers that apply LLMs to build systems across different domains, including the total that use Chain-of-Thought prompting.}
  \label{fig:domain}
\end{figure*}

\subsubsection{Interactions Within System}
\label{sec:hci-exec-sys}
Our survey indicates that 178 (69\%) LLM-related papers apply LLMs to build user-interacting systems across various domains (Figure \ref{fig:domain}).
Of these works, about half (N=82, 46\%) build multi-LLM systems, involving an orchestration of several tasks that are performed by multiple LLMs, independently or collaboratively, which are sometimes supported by other components such as external classifiers or algorithmic workflows.
Although the LLMs within a system may each be carrying out separate, simpler reasoning tasks, these underlying reasoning processes are not always independently interpretable or articulated in the system's final outcome. Instead, they come together and take on different forms, shaped by the kinds of responses users seek \cite{Zhang2023AppAgentMA,choi2025proxona,ma2025dbox,Riche2025AIInstrumentsEP}.
For example, \citet{Ge2025GenComUIEG} develop a multi-LLM system to introduce visual aids such as map annotations and path indicators on an assistant robot's screen to support human-robot interaction; though LLMs here perform CoT reasoning underneath to interpret user intent, understand requirements, and transform them to code, the final output is a multimodal rendering of user instructions.
In this context, we observe that prior works on multi-LLM systems often interpret the execution of LLM reasoning in relation to the \textit{interactions} among different components/agents within the system, instead of only through fine-grained stepwise instructions and domain knowledge.
    
For example, to generate real-time interactive 3D world models from users' natural language prompts, \citet{DeLaTorre2023LLMRRP} proposed a linear stack of LLMs tasked with planning, scene analyzing, code and generation, and checking. While these operations are performed in sequence, each requiring a different form of reasoning (and many utilizing CoT-like prompts), the authors also interpret the coordinated outcome of their system as LLM's \textit{``capability of spatial reasoning and world understanding''}.
Similarly, \citet{Earle2024DreamGardenAD} use LLMs to transform a high-level user prompt into a broad plan, which is then recursively broken down into sub-goals. While the individual LLMs engage in localized, small-scale planning-oriented reasoning, the LLM-system's overall reasoning is expressed through its ability in hierarchical planning and the generation of concrete coding tasks.
In addition to such planning-based modules in several works \cite{Zhang2025FrictionDW,pu2025ideasynth}, the emergent reasoning behavior in many studies also involve other sub-tasks such as seeking clarifications \cite{chen2025dango}, rationalizing explanations \cite{Ma2024TowardsHD}, generating sequential story plots \cite{Lu2025WhatELSESN}, and domain-based scaffolding questions and follow-ups \cite{Liu2025ScaffoldedTA,Skeggs2025MicronarrativesAS}, among others.
An example could be \cite{Ma2024TowardsHD} where individual LLMs take on the role of analyzing a user's perspective on a given topic, providing thoughtful and critical responses, and evaluating human arguments on logical consistency. Here, the reasoning of the LLM system as a whole is oriented toward supporting human–AI deliberative decision-making.
Collectively, these studies highlight that LM reasoning is many times carried out through the interactions among multiple components of the HCI system in which LLMs are embedded, rather than being confined to prompting strategies alone. 

\subsection{Evaluation of Reasoning} 
\label{sec:hci-eval}
While the reasoning behavior of LLMs is often not the primary object of evaluation in HCI, we identify two nuanced intertwined characteristics in how LLM reasoning is assessed: an emphasis on users' subjective utility and the obscurity in the evaluation process itself.

\subsubsection{Subjective Utility-Based Validation}
\label{sec:hci-eval-subj}
In NLP and ML, to compare and contrast the reasoning abilities of LLMs, their responses are typically evaluated using quantitative metrics on benchmark datasets covering different tasks such as commonsense, logical, mathematical, causal, and coding. 
In contrast to evaluating the correctness of reasoning outcomes in these seemingly deterministic tasks, since the end tasks in HCI are user-centric and often subjective, the prior works at CHI primarily focus on evaluating how \textit{useful} the LLMs' reasoning behaviors and outcomes are to the target users.
For example, \citet{goldi2025efficient} evaluate the reasoning goodness of their LLM-based coaching agents, not in terms of outcome correctness, but along domain-specific axes such as user satisfaction, interaction enjoyment, and expectation confirmation.
To examine the role of CoT-based LLM reasoning in facilitating MATLAB learning, \citet{Rogers2025PlayingDT} conducts user studies investigating several questions around how the LLM's learning-by-strategy approach shapes students' self-efficacy and their attitudes toward programming. 
Some other illustrative examples can be seen in areas such as human-AI collaborative decision-making \cite{Ma2024TowardsHD,Pafla2024UnravelingTD}, user interface development \cite{kretzer2025closing,zhang2025appagent}, mental and physical health \cite{Choi2024PrivateYS,Sharma2023FacilitatingSM}, learning assistance \cite{Lawley2023VALIT,Pu2025AssistanceOD}, and idea generation \cite{Lin2023JigsawSD,hou2024c2ideas}.

On the other hand, there are a few works that measure the correctness of LLMs' reasoning capabilities in providing specific responses where ground truths are unambiguous \cite{huang2024plantography,Li2024OmniActionsPD,ding2024leveraging}.
For instance, \cite{ding2024leveraging} evaluate how their incremental reasoning strategy results in accurate identification of cause-effect pairs that are objectively determined in social media posts; however, this serves only as an intermediate measure, with the primary focus on the practical utility of these pairs and their summaries for understanding public attitudes and health decision-making.
In almost all of these studies, the focus is not on comparing the reasoning capabilities of multiple models or demonstrating how their methods improve LLM reasoning. In other words, LLM reasoning itself is not examined in a controlled environment to perform a task and then evaluated (like in NLP/ML); rather, it is assessed indirectly through its practical utility in supporting broader objectives

\subsubsection{Latent Validation} 
\label{sec:hci-eval-latent}
As discussed above, since LLM reasoning is primarily interpreted as playing an instrumental role in prior works, whose main objective is to build systems and frameworks, its evaluation is consequently not evident in some studies.
We particularly find such implicit cases where LLM reasoning serves as a motivator for applying LLM, and/or when LLMs' reasoning behavior is imagined and expressed in ways that diverge from commonly observed forms.
Consider the study by \citet{Kang2025BioSparkBA} that builds an LLM-based application to support creative idea generation by drawing analogies from biological organisms and systems.
They use breadth- and depth-focused approaches to expand seed ideas, resembling the step-by-step generation of reasoning trajectory in recent LLMs; yet the reasoning capabilities of LLMs are not explicitly presented as an object of their evaluation.
In another related study, \citet{Jones2023EmbodyingTA} evaluate the creative reasoning \cite{hintikka1996creativity} side of LLMs by prompting GPT-3 to generate human-like rules of engagement for performance artists in diverse contexts.
These types of works extend beyond Creativity and span across domains, where they develop LLM applications for different forms of reasoning, but their evaluation largely remains latent \cite{Lin2025SeekingIT,Liu2023HowAP,wu2024mindshift,Deva2025KyaFP}.
This scenario stands in contrast to the cases examined at the end of section \ref{sec:hci-frame-motiv}, where the reasoning capabilities of LLMs were invoked as a rationale for their application, even though they were not essential.

\begin{table*}[]
\centering
\resizebox{\textwidth}{!}{%
\begin{tabular}{|>{\columncolor{gray!7}}l|>{\columncolor{gray!7}}l|>{\columncolor{blue!5}}l|>{\columncolor{red!5}}l|}
\hline
\textbf{\cellcolor{gray!20} Dimension} & 
\textbf{\cellcolor{gray!20} Foundations in NLP/ML} & 
\textbf{\cellcolor{blue!15} Abstractions in HCI} & 
\textbf{\cellcolor{red!15} Consequences for HCI} \\ \hline
\textbf{Framing} & \begin{tabular}[c]{@{}l@{}} \\ [-2mm] Structured process: \\ Organization of atomic steps \\leading to a terminal state \\ [2mm]\end{tabular} 
& \begin{tabular}[c]{@{}l@{}} \\ [-2mm] \textbullet\ Prior benchmark results and \\ studies serve as motivators \\ for applying LLMs \\ \\ \textbullet\ Imagined as linear chain of \\ sub-tasks progressing \\ sequentially \\ [2mm]\end{tabular} 
& \begin{tabular}[c]{@{}l@{}} \\ [-2mm] \textbullet\ Steps loaded with sub-tasks \\ and abstract guides \\  \\ \textbullet\ Unexplored organization of \\ steps or sub-tasks, such as \\ tree-of-thought or\\ graph-of-thought \\ [2mm] \end{tabular} \\ \hline

\textbf{Execution} & \begin{tabular}[c]{@{}l@{}} \\ [-2mm] \textbullet\ Training to imitate \\ reasoning-like data \\ \\ \textbullet\ Training to imitate \\ step-by-step reasoning \\ through reward signals \\ [2mm] \end{tabular} 
& \begin{tabular}[c]{@{}l@{}} \\ [-2mm] \textbullet\ Fine-grained prompting, \\ through domain knowledge \\ and personas, as the gold \\ standard to elicit reasoning \\ \\ \textbullet\ Through interactions among \\ different system components \\ [2mm] \end{tabular} 
& \begin{tabular}[c]{@{}l@{}} \\ [-2mm] \textbullet\ Ignoring the limitations of \\ CoT, especially in terms of \\ influence of demonstrations \\ \\ \textbullet\ Overlooking the benefits  \\ of SFT for tasks whose datasets \\ may be under-represented \\ \\ \textbullet\ Lack of contribution of \\ domain-specific \\ reasoning-like data \\  \\ \textbullet\ Wide use of closed- \\ source models \\ \\ \textbullet\ Ignoring the design of \\ underlying reward models \\that significantly influence \\ reasoning step generation  \\ [2mm] \end{tabular} \\ \hline

\textbf{Evaluation} & \begin{tabular}[c]{@{}l@{}} \\ [-2mm] \textbullet\ Validating response in its \\ its entirety only \\ \\ \textbullet\ Validating every step and \\ how it leads to terminal state \\ [2mm] \end{tabular} 
& \begin{tabular}[c]{@{}l@{}}  Subjective utility-based \\ validation  \end{tabular} 
& \begin{tabular}[c]{@{}l@{}} \\ [-2mm] \textbullet\ Evaluation specific to \\ reasoning remains latent \\ in some applications\\ \\ \textbullet\ Over-emphasis on plausibility- \\based evaluation, ignoring \\ model faithfulness \\ \\ \textbullet\ Less consideration towards \\ individual steps or underlying \\ process due to an over-focus \\ on the complete response's utility \\ [2mm] \end{tabular} \\ \hline
\end{tabular}
}
\caption{Summary of our key findings. For each dimension of our research question, we map the foundational elements in NLP/ML that are abstracted away in HCI research. The last column describes the critical consequences of such abstractions.}
\label{tab:findings}
\end{table*}

\section{HCI's Abstraction of LLM Reasoning}
\label{sec:hci-abs}
In light of our survey on how HCI approaches LLM reasoning, we now turn to the NLP and ML literature for contextualization. 
Our goal in this section is two-fold: (1) to situate HCI's engagement with LLM reasoning within the broader discourse in NLP/ML and discuss its structural and methodological abstractions from the foundational elements of reasoning, and (2) to identify potential opportunities for future works in HCI. 
While discussing the foundations in NLP/ML, we deliberately set aside the mathematical formalism and provide a conceptual account that preserves the core ideas, making them broadly accessible to HCI researchers from diverse backgrounds. 

\subsection{Framing of Reasoning}
\label{sec:hci-frame-abs}
LLM Reasoning is defined in multiple ways in NLP/ML---based on the actions an LLM performs or by excluding tasks it does not---often by alluding to its usage in different fields such as philosophy, medicine, or physics (see \cite{yu2024natural,sun2025survey} for a survey of definitions). 
Here, we zoom out to consider how reasoning is ``framed'' or positioned in order to bring out the essential components that underpin the rhetorical aims of many invoked definitions of reasoning.
We find that most LLM reasoning behaviors in NLP/ML are framed as \textit{structured} processes with three essential components: (i) \textbf{reasoning steps}, the atomic units of inference or computation; (ii) the \textbf{organization of steps}, which specifies how these units are connected to serve different purposes; and (iii) the \textbf{terminal state}, the desired outcome toward which the process is directed.
However, in the \textit{motivators} framing adopted in many HCI studies (section \ref{sec:hci-frame-motiv}), LLMs' reasoning capabilities are often considered as \textit{givens} in the process of applying LLMs to build systems, which in turn, abstracts out these structural elements of reasoning.

As discussed in section \ref{sec:hci-frame-seq}, reasoning steps in HCI are often semantically coherent segments of texts packed with sub-tasks and abstract instructions.
This complexity is not inherently problematic; indeed, several works in NLP/ML have successfully incorporated complex reasoning steps---the most notable example being the DeepSeek-R1 training methods \cite{guo2025deepseek}, where each reasoning step includes a complete response\footnote{While it was discussed in the context of training DeepSeek-R1 \cite{guo2025deepseek}, what we emphasize here is that complex reasoning steps are acceptable and useful for many applications.}.
However, under the veil of sequential CoT-style progression adopted in HCI, each reasoning step is often uncritically interpreted as an atomic unit in the solution space.
While it is common to include formatting and stylistic instructions in steps, if the steps are, however, overloaded with sub-tasks and require subjective interpretation, each step may itself be expanded in non-linear pathways by the LLM, exploring multiple possible approaches, many of which are overlooked in prior HCI works.
In many instances, a reasoning step is used as a proxy for an underlying sub-task, though this distinction is rarely taken into account in framing LLM reasoning.
We do not suggest breaking down the steps to the token level; on the other hand, we want to emphasize that, while choosing a step granularity aligned with an objective, the researcher must ensure that each step is well-scoped for their task and investigate the effects of different levels of abstraction.

The overreliance on CoT has also made HCI overlook other non-linear modes of expanding reasoning steps.
An increasingly explored alternative is Tree-of-Thoughts (ToT) \cite{yao2023tree}, where multiple branches of ``thoughts'' are explored like a \textit{tree}, and better solutions are selected at each level until a solution is reached.
In some works, multiple independent linear chains of thoughts are explored in parallel \cite{wang2022self, ning2023skeleton}.
The most generic or arbitrary form of organizing different reasoning steps is through \textit{graphs}, where multiple steps can also merge into one another \cite{besta2024graph}. 
While the advantage of employing a tree structure lies in exploring multiple diverse reasoning paths, for graphs, it lies in producing better paths/outcomes by combining different paths \cite{besta2025demystifying}.
When a reasoning step is prescribed in a complicated way, different sub-tasks within that step may be explored in graphical modes where multiple contextual information could be integrated to approach a sub-task.

We find a few works that implicitly use a non-CoT organization of steps---such as harnessing multi-turn conversations \cite{jin2024teach,Epperson2025InteractiveDA}, multiple parallel reasoning paths \cite{Lu2025WhatELSESN,cheng2024relic}, and graph expansion \cite{qin2025toward,Riche2025AIInstrumentsEP}---though they did not explicitly present as such.
For instance, in \cite{Riche2025AIInstrumentsEP}, the vertical and horizontal exploration of image generation from text prompts resembles and could benefit from an explicit graph-like reasoning frame.
Being aware of these nuanced elements of reasoning framing would support HCI researchers in understanding and being informed of how LLMs may expand their inputs.
We refer readers to the recent work of \citet{besta2025demystifying}, which provides illustrative examples of diverse reasoning frames across multiple domains, such as creative writing, puzzles, and games.




\subsection{Execution of Reasoning}
\label{sec:hci-exec-abs}
An LLM can execute a task---while adopting a reasoning frame---either by being explicitly trained on it or by inference scaling. 
We discussed in section \ref{sec:hci-exec-gold} that inference scaling is widely adopted in HCI, relying on a range of prompting techniques, with or without domain knowledge, and in many cases involving the coordination of multiple LLMs or external tools.
Of particular relevance in this context is the practice of including domain-specific demonstrations or sample input-output pairs in the prompt for the LLMs to \textit{learn} from, a technique commonly referred to as in-context learning \cite{dong2022survey,zhou2022teaching}.
While this has been advanced as an alternative in many HCI papers we reviewed \cite{zamfirescu2025beyond,yang2025geniewizard,hou2024c2ideas}\footnote{This is also a commonly followed practice in NLP/ML, though it is being increasingly critiqued recently \cite{stechly2024chain,zheng2025curse}.}, recent research in NLP/ML discuss several challenges in understanding the required level of specificity in the demonstrations or the similarity to the target problem, even for deterministic simple planning problems \cite{stechly2024chain,zhang2025prompt,zheng2025curse}.
In general, several works highlight how LLMs struggle to infer necessary patterns from demonstrations and generalize to diverse contexts.  
On the other hand, HCI is flooded with subjective human-centric tasks, where how and why prompt engineering works are yet to be theoretically explored.

While HCI literature primarily relies on user studies to validate LLM responses (section \ref{sec:hci-eval-subj}), due to the reasons discussed above, it becomes important to examine the process an LLM potentially followed from its training to generate the reasoning-like response at inference, in order to accurately understand and trust its outcome.
This not only requires prompt engineering and tracing of the self-reported steps it followed for consistency checks (like in \cite{cheng2024relic}), but also an investigation of its underlying training methods, reasoning-like datasets used for training, and consistency with the reasoning framings adopted during training.
By training to reason, prior works in NLP/ML often refer to the \textit{fine-tuning} process, where the objective is to add reasoning abilities to LLMs on top of their next-token prediction abilities; in other words, the goal is to generate reasoning-like responses.
Below, we will discuss two broad classes of training (or fine-tuning) that remain almost entirely abstracted out in HCI.
    

\subsubsection{Training From Reasoning-Like Data}
\label{sec:hci-exec-abs-train}
The first approach shares the same objective as pretraining LLMs (for generic text generation) but involves fine-tuning on text datasets tailored to specific reasoning tasks. 
Each dataset instance typically pairs an instruction or question with a response that follows the reasoning frame required for the task. 
Human annotation is the most straightforward way to curate such datasets.
In mathematical reasoning, for instance, an annotated instance may consist of either a linear sequence of steps leading to the solution (a chain structure) or a branching exploration of alternatives at each step, with progression determined by specific criteria (a tree structure) \cite{lightman2023let,besta2025demystifying}.
However, this process is resource-intensive and often infeasible for many tasks. Consequently, most prior works in NLP/ML rely on external tools or LLMs themselves to generate intermediate reasoning steps, either partially or fully\footnote{Some examples of external tools are WolframAlpha for rewarding mathematical reasoning and PerspectiveAPI for toxicity reasoning}.

In this process, model parameters are updated such that the next-token predictions reflect a reasoning frame consistent with the training data and prompt instructions, though how this is balanced is still an open question \cite{parthasarathy2024ultimate,lyu2024keeping}. 
Such trained models can also be prompted to generate one semantically coherent step at a time without explicit instruction of what constitutes a step, if special tokens are inserted between steps during training \cite{Besta2025ReasoningLM}.
These special tokens act as structural markers, signaling where a reasoning step begins and ends. During training, the model learns to interpret them as delimiters that separate steps, rather than as part of the response content
This entire training approach is called Supervised Fine-Tuning (SFT).

While SFT may not be necessary for all tasks---particularly those requiring only generic text generation---it has been shown to improve LLM reasoning when domain-specific knowledge is underrepresented in pretraining.
Despite its central role, HCI studies rarely examine SFT training datasets to assess whether their desired forms of reasoning can potentially emerge when LLMs are prompted.
A particular focus also needs to be given to how reasoning steps are constructed and organized during training.
If reasoning frames (which may differ across problems within the same task) accurately reflect how reasoning should be approached for the target task, they increase the likelihood that the model will internalize expected coherence within each step, recognize connections across steps, and learn how to combine them.
Conversely, misaligned or poorly scoped reasoning frames may limit the model's ability to approach the problem effectively. In this sense, SFT does not simply ``improve reasoning'' in general but conditions the style and granularity of reasoning that the model internalizes.

Recent findings further complicate this picture. Even after SFT, prompting techniques CoT may fail, partly due to asymmetries between implicit learning from fine-tuning and explicit guidance from demonstrations in the prompt \cite{zhang2025prompt,stechly2024chain,zheng2025curse}. This tension underscores the fragility of reasoning alignment: gains from SFT may be undermined if inference-time prompting introduces conflicting or mismatched expectations. Such risks become particularly acute in high-stakes domains such as healthcare, where misaligned reasoning could have disproportionate consequences.
While some prior works have explored supervised fine-tuning (for e.g., \cite{jones2025artificial,huang2024plantography}), its adoption remains limited---only 9\% of studies in our survey applied it---and in certain cases, SFT was even conflated with inference scaling \cite{szymanski2024integrating}.

Another area where HCI can make a significant contribution is the creation of domain-specific datasets for SFT. In NLP and ML, many studies leverage LLMs' in-context learning abilities to generate such datasets, often employing human-in-the-loop setups to ensure quality (for e.g., \cite{liu2022wanli,lightman2023let}).
Although this has not traditionally been a central focus in HCI, developing well-structured datasets can support replication and improvement of existing systems, and enhance LLMs' underlying reasoning capabilities for similar tasks.
For example, consider \citet{Ko2023NaturalLD} who use CoT prompting to create diverse natural language datasets for advancing LLM reasoning in data visualization research.
However, the challenge here is that datasets resembling reasoning are not always straightforward to construct. In many domains, accurate reasoning traces may be difficult or impossible to capture, and real-world data often contains mis-reasonings, corrections, or false starts as part of the task. For example, in conversational HCI tasks, such as in providing emotional support \cite{zheng2025customizing}, the natural reasoning process is nonlinear and may contain errors. Training on such data through SFT risks instilling flawed reasoning patterns, since the model learns to mimic both correct and incorrect steps
Therefore, it is important to distinguish high-quality reasoning steps from flawed ones during dataset construction. The next section discusses approaches to achieve this in more detail.

\subsubsection{Training Using Reward Signals}
\label{sec:hci-exec-abs-infer}

The second method for training aims to refine an existing reasoning-generating LLM further (like the one trained using SFT), which we will call the reasoner LLM here, based on external signals that \textit{reward} better reasoning steps over poor ones.
The reasoning steps can also be generated by multiple LLMs, like in the case of debate-like reasoning, and the rewarding signals can be provided by external tools, APIs, or other LLMs. This process is referred to as reinforcement-learning (RL) based fine-tuning.

The core idea of this paradigm is that, in each iteration, the reasoner LLM is prompted to generate multiple reasoning steps---often structured as a tree to explore alternative paths---for a given problem\footnote{Note that, unlike SFT, the fine-tuning here is not on given problem-response pairs.}.
A reasoning step can also be a complete solution, as in the case of DeepSeek-R1 \cite{guo2025deepseek}, where multiple complete solutions are explored at each level, and refinement of entire solutions happens. 
A reward model then evaluates these steps, selecting promising paths while pruning or refining others, with the reasoner's parameters updated from reward signals.
This process may involve step refinement, backtracking, or solution revision, and continues until a stopping criterion is met.
The rewards are often numerical values that quantify the extent to which a step will lead to the correct solution.
In sum, while SFT relies on curated reasoning-like datasets to predict the next sequence of tokens, the RL-based paradigm finetunes LLMs through external rewards in an explore–exploit setup.
While the reasoner itself does not require annotated data, a reward model, which can be an LLM, must be trained to reliably assess reasoning steps.

As most off-the-shelf LLMs, like GPT-series used in almost all HCI studies, undergo RL-based training \cite{jaech2024openai}, it becomes important to examine and understand what and how reward models were used during this training phase.
If the reward models are outcome-oriented, discounting the nuanced signals for distinguishing high-quality intermediate steps from flawed ones, then the reasoner LLMs may not follow the desired reasoning process at inference.
For instance, the reward models used for training in most commonly used LLMs often revolve around high-level overall quality and the correctness of the final conclusion, and focus largely on deterministic tasks \cite{uesato2022solving,havrilla2024glore}.
While such models may be producing desired responses due to extensive prompting effort and user studies may reinforce this, it is important to cautiously set the expectations and imaginations by examining these training pipelines, as they can be particularly critical for high-risk applications such as mental health counseling, seeking medical clarifications, or mimicking student/teacher behavior in children's education.
This introduces an added layer of complexity: the reward function must be carefully designed to reflect the nuanced goals of the target task. In some cases, heuristics suffice---for example, in math tasks, correctness of the final solution can serve as a reward signal \cite{lightman2023let}. In other scenarios, external tools (e.g., the Perspective API), domain-specific rules, or auxiliary models may be necessary to formalize reward structures.
In our survey, we find that only \citet{wang2025exploring} explicitly engage with RL-based fine-tuning, taking into account dynamic user behavior and environmental context to balance familiar and novel experiences in order to motivate sleeping behavior change.
This limited attention highlights a broader gap: although many studies generate diverse user feedback data, which could potentially contain implicit penalties and rewards of different reasoning steps, these signals are rarely systematically exploited to improve reasoning.


\subsection{Evaluation of Reasoning}
\label{sec:hci-eval-abs}
In HCI, while the term ``evaluation'' is often associated with judging the quality of LLM responses for specific tasks at inference time, the reasoning behavior of LLMs is also evaluated during the training stage, which is rarely considered.
Across both training and inference, evaluation can be broadly understood as occurring at two distinct levels, a dimension that is largely overlooked.

\subsubsection{Validation As a Whole}
\label{sec:hci-eval-abs-whole}
The most common and straightforward approach to evaluating LLM reasoning is to manually assess a reasoning response in its \textit{entirety}, as is typically followed in HCI user studies. 
This approach emphasizes the assessment of \textit{plausibility} of the reasoning process to humans, but often neglects to understand whether the LLM has in fact faithfully executed the reasoning steps it presents \cite{agarwal2024faithfulness,paul2024making}.
Model faithfulness is an ongoing challenge in ML/NLP, particularly in subjective reasoning tasks such as explaining why a text might be toxic or hateful \cite{mothilal2025human}.
In particular, several works in NLP/ML have highlighted the unfaithfulness of outcomes derived through CoT reasoning \cite{lyu2023faithful,creswell2022faithful,yeo2024interpretable}.

However, HCI, largely being user-facing, hardly pays attention to how reliable the LLM reasoning outcomes are while focusing on their subjective utility (see section \ref{sec:hci-eval-subj}), potentially misleading the end users in many cases.
\citet{cheng2024relic} is the only work in our corpus that explicitly engages with the task of evaluating the faithfulness of LLM responses (though for a very specific Q/A scenario) by generating multiple responses to the same query and checking for consistency across them. 
Though not positioned explicitly in the context of LLM reasoning, studies like these can contribute to understanding the underlying reasoning of LLMs in producing subjective outcomes.
While recent works in ML/NLP emphasize moving beyond consistency-based measures toward examining model architecture for faithfulness \cite{zhang2025prompt}, investigating domain-specific self-consistency of LLM responses---while remaining well within the scope of HCI research---remains a necessary, if not sufficient, condition for reliability of LLM responses.

Another important form of evaluating LLM reasoning that largely remains opaque to HCI is the RL-based methods discussed in section \ref{sec:hci-exec-abs-infer}, where external rewards or feedback are used to continuously explore and finetune the LLMs towards better intermediate steps.
In many cases, rather than evaluating every step in a reasoning tree, multiple complete responses are evaluated and ranked using reward signals.
This approach falls under the evaluation-as-a-whole paradigm, in which one or more candidates are selected from a set of alternatives. 
In NLP/ML, this RL-based training strategy is specifically referred to as preference learning or alignment tuning, reflecting the process of favoring one candidate over others. When these preferences are provided by human annotators, the method is commonly termed as reinforcement learning from human feedback (RLHF) \cite{christiano2017deep}.
\citet{chakrabarty2025can} is the only study in our review that directly addresses preference learning by proposing a method to compare human responses with LLM-generated outputs that have been edited according to domain-specific criteria. They generate over a thousand finely edited samples, enabling RL-based tuning of LLMs across a range of writing tasks, from fiction and food writing to online advice. 
While collecting and analyzing user preference data through different modes is a well-established focus in HCI (for e.g., see \cite{qin2025toward,Chakrabarty2024ArtOrArtifice,aoyagui2025matter}), it has rarely been leveraged to support LLM training.
Following \cite{chakrabarty2025can}, there is a clear avenue for HCI to contribute domain-specific preference datasets for fine-tuning LLMs for specialized tasks.

\subsubsection{Validation In Parts}
\label{sec:hci-exec-abs-infer}
The second broad approach is to evaluate one or more parts of the reasoning process, particularly the intermediate steps or the terminal step(s).
We note two key abstractions here.
The first is the tendency to cite prior HCI studies as evidence of LLMs' reasoning capabilities. While such studies may suggest reasoning, it is important to consider that most of the reasoning validation in HCI focuses on subjective utility (section \ref{sec:hci-eval-subj}) and so may not generalize to new contexts.
Secondly, many HCI studies unreflectively rely on LLMs' reported performance on benchmark datasets.
Although the tasks may require reasoning, most of the benchmark datasets contain only the final response to a given problem \cite{Sun2023ASO,Besta2025ReasoningLM}.
The assumption here is that, due to supervised fine-tuning on a large scale of reasoning-like datasets and reward-based supervision, the LLM is presumed to encode a reasoning process in generating the final response.
However, this depends on the availability of ground truth labels and how aligned the data curation process is with the target task.

A similar approach is also prevalent during training phases where outcome-based reward models (ORMs) are used in RL-based fine-tuning.
ORMs assign rewards only to the terminal state of a reasoning trajectory, disregarding the quality of intermediate steps \cite{uesato2022solving}.
Since evaluating every reasoning step is often infeasible for many tasks, prior works in NLP/ML often adopt this approach, where they expand a reasoning trajectory and, if the terminal state is desirable, propagate a positive reward backward to all steps, irrespective of the quality of an intermediate step.
On the other hand, in Process-based Reward models (PRMs), rewards are also assigned at all intermediate steps depending on how good they are, sometimes independently, and at other times, in relation to the terminal state \cite{lightman2023let,uesato2022solving}.
While PRMs are resource-intensive due to the need to evaluate every step, ORMs risk reinforcing incorrect intermediate reasoning towards the correct solution.
For HCI research, this distinction underscores the importance of scrutinizing how reward models shape the evaluation of reasoning, especially as RL-based fine-tuning underlies many recent off-the-shelf LLMs.

\section{Discussion}
\label{sec:checklist}

We have discussed how HCI research abstracts out several structural and methodological foundations of LLM reasoning in the previous two sections.
While Table \ref{tab:findings} summarizes these abstractions, the breadth of contributions in this area can be overwhelming and leave HCI practitioners uncertain about where to begin.
To navigate this landscape, we build on our review of both HCI and NLP to develop a conceptual tool that supports HCI practitioners in critically engaging with reasoning when applying LLMs to their tasks, and remain attentive to the abstractions they may adopt or introduce in the process.

A substantial body of work in HCI, NLP, and related fields has developed tools and frameworks to help AI practitioners critically engage with different stages of the AI pipeline, from data collection to post-hoc fairness analysis \cite{taylor2015data,bhat2023aspirations,diaz2022crowdworksheets,pushkarna2022data,Jarrahi2022ThePO}. 
These efforts encompass both general-purpose and domain-specific approaches; notable examples include model cards \cite{mitchell2019model} and datasheets \cite{gebru2021datasheets}, which have become widely adopted and are frequently released alongside LLMs.
Despite these advances, a persistent concern across these tools and frameworks is the lack of clear guidance for practitioners on how to interpret and respond to some of the critical prompts for analysis and documentation \cite{wong2023seeing,kommiya2024towards}.
The prompts are often too generic and ambiguous, leaving practitioners uncertain about what kind of input is expected.
As a result, practitioners often rely on implicit assumptions or offer surface-level responses, which limit opportunities for deeper reflection and diminish the effectiveness of these tools \cite{Liang2024SystematicAO,KommiyaMothilal2025TalkingAT}.

By drawing inspiration and learning lessons from these works, we propose a conceptual tool for LLM reasoning consisting of a structured set of reflection prompts, guided by two design choices: (1) tailoring prompts to specific steps that practitioners might---or should---consider when reasoning about LLMs in relation to their task, (2) positioning the tool as an argument pad rather than a checklist, encouraging reflective and justificatory responses rather than perfunctory completion.
Specifically, by argument, we mean that each response to a reflection prompt should articulate a stance---which may be indeterminate or multifaceted---accompanied by reasons or evidence that justify that stance. 
While practitioners may already engage in such reasoning informally with existing tools, making this process explicit in the context of LLM reasoning offers two advantages.
First, it sets up the intention for the argument subject to \textit{reason} about LLMs' reasoning abilities and externalizes the deliberative process before using them.
Second, framing responses as arguments also opens up a space for debates from others---for critiquing or evaluating---which is an interesting area to explore in the future.

While we have drafted the reflection prompts based on what we inferred from our reviews across NLP/ML and HCI, we recognize this may not be exhaustive, and some HCI works might need to consider different scenarios, such as prompts specific to multimodal reasoning. 
To support ongoing development and community input, we will open-source the prompts as a living document, allowing others to contribute and refine approaches to LLM reasoning in HCI.
Now, we introduce our reflection prompts by grouping them into four themes, reflecting our reviews: reasoning identification, framing, execution, and evaluation. 
Though we present them in sequence, they inform each other, and the prompts are intended to be applied iteratively and referenced across themes.

\subsection{Reasoning Identification}
We highlight in sections \ref{sec:hci-frame-motiv} and \ref{sec:hci-eval-latent} that HCI practitioners may conflate or misinterpret whether LLMs are engaging in reasoning in different contexts.
To address this, explicitly identifying if some form of reasoning is involved in the use of LLMs provides a good starting point for reflective use.
Our prompts below are designed to facilitate this process by focusing on the presence of RLMs in the pipeline, the expectations for the target task, and the subjective interpretation of the LLM response.
In particular, we recognize that an LLM may be reasoning even when the task does not require it, and conversely, the response may appear devoid of explicit reasoning while the model is implicitly doing it.

\begin{mybox}[Reasoning Identification]
\begin{enumerate}
    \item Does your system consist of an explicitly trained reasoning language model? Explain why you say that it is (or it is not) a reasoning model?
    \item Does the task your system is supposed to do involve some form of reasoning as per your interpretation? Justify your interpretation.
	\item Does the expected response (at inference, before or after reasoning finetuning) involve reasoning as per your interpretation? Justify your interpretation.
\end{enumerate}
\end{mybox}

\subsection{Reasoning Framing}
Once the presence of LLM reasoning has been established, it is important to set the framing right since this informs how reasoning is both implemented and evaluated later.
Importantly, practitioners may only identify reasoning after reflecting on the framing itself, underscoring the interdependence of identification and framing.
Below, we structure this section parallel to the identification process, but with greater granularity.
In some cases, expected reasoning from LLMs can be framed directly based on task requirements. 
However, in others, reasoning, identification, and framing can be approached only after seeing LLM responses for a task. This might particularly be the case for domains such as Design or Creativity, where, as discussed in section \ref{sec:hci-eval-latent}, LLMs' reasoning behavior may be latent or split across multiple prompts.
Our reflection prompts are thus designed to critically engage with reasoning frames, particularly the uncritical adoption of linear CoT prompting in practice (section \ref{sec:hci-frame-seq}).

\begin{mybox}[Reasoning Framing]

\noindent \textbf{Reasoning process required from LLMs (which may be reflected in the response to different degrees)}:
\begin{enumerate}
    \item Is the required reasoning process (from the LLMs) separated into multiple parts or sub-tasks? If so, respond to all the following questions that apply.
    \begin{enumerate}
		\item Is the reasoning process structured in an orderly step-by-step manner? Justify why this is necessary.
		\item Is the reasoning process structured in stages or phases (where each stage may have its own sub-task) with a degree of order between them? Justify why this is necessary.
		\item List down all possible interpretations that a reasoning step or stage or phase can refer to in your application. Justify your interpretations.
		\item Does a step or stage or phase include multiple reasoning-related sub-tasks? If so, explain why they are necessary and how they are related to LLM reasoning.
		\item If the expected reasoning is distributed across multiple responses, then how are individual steps or stages or phases interpreted?
		\item How are individual steps or stages or phases organized and interacted with one another? Do the steps branch and/or merge? Justify why your desired mode of organization is necessary.
    \end{enumerate}
    \item Does the terminal state of your reasoning task have one or more than one possible conclusions? Explain why.
	\item Can each of the possible conclusion of your reasoning task be objectively understood or are they open to subjective interpretations? Justify the objectiveness of your preferred interpretation.
\end{enumerate}

\noindent \textbf{Reasoning behavior interpreted from LLM response:}
\begin{enumerate}
    \item Is your interpreted reasoning behavior at inference expected to follow some structure or form? Justify why a structure is (or is not) required and explain the level of explicitness in the structure, if applicable.
	\item Can the reasoning behavior be structured in ways other than by separating it into multiple parts or sub-tasks? If so, explain why this might be preferred.
    \item To what extent is the reasoning behavior localized to individual responses? Is it (also) distributed across multiple responses, like in a multi-turn conversation? Justify the distributedness.
\end{enumerate}
\end{mybox}

\subsection{Reasoning Execution}
As discussed in section \ref{sec:hci-exec-gold}, inference scaling has emerged as the gold standard for applying LLMs in HCI, extending effectively to reasoning tasks as well. However, prior work has rarely explained why fine-tuning approaches are often dismissed or left unexplored. This section introduces prompts intended to encourage practitioners to critically reflect on the dominance of inference scaling and to consider alternative approaches, where appropriate. Even when fine-tuning is not adopted---due to practical, technical, or resource constraints---the prompts here aim to foster awareness of training processes of off-the-shelf models, and their inherent limits and capabilities. Finally, as in other sections, we emphasize that the framing of tasks can influence how reasoning is carried out, just as the constraints of execution can, in turn, shape the framing itself.

\begin{mybox}[Reasoning Execution]
\noindent \textbf{Prompting strategies (inference scaling):}
\begin{enumerate}
    \item Does your task rely solely on prompting techniques? If so, what strengths and capabilities of LLMs convince you that this is sufficient?
	\item If you considered some form for finetuning LLMs but did not implement it, justify why.
	\item Did you include any domain knowledge in your prompt(s)? If not, justify why this is not required or explain how you interpret LLMs can incorporate the necessary knowledge.
	\item Were your final prompt(s) evolved over several iterations? Explain what changes were necessary and why.
	\item How did you choose in-context examples or demonstrations for your task? Justify why these samples will help LLMs to respond better.
    \item Do you explicitly introduce the reasoning frame, required for your task, in the prompt? Justify your stance.
\end{enumerate}

\noindent \textbf{If (you think) your system is finetuned on reasoning-like datasets relevant for your task (supervised finetuning), then respond to all the following questions that apply:}
\begin{enumerate}
    \item To what extent do the contents of the finetuning datasets reflect the kinds of responses your system is expected to produce at inference? 
	\item Are the ways reasoning is structured (steps, stages, or phases) and organized (in chains, trees, or graphs) in finetuning datasets similar to how your system is expected to respond at inference?
	\item If the finetuning datasets are not specifically developed for your task, justify why task-specific datasets are not used.
	\item Does your finetuning datasets contain flawed reasoning patterns (as per your task)? If so, explain how this is (or is not) a problem; if not, justify your stance.
    \item Does your system's performance on tasks other than (or somewhat related to) yours change due to the finetuning? If this aspect is insignificant, explain why.
\end{enumerate}

\noindent \textbf{If (you think) your system is finetuned using external reward signals (reinforcement learning), then respond to all the following questions that apply:}
\begin{enumerate}
    \item What factors informed your choice of reward model, and why?
	\item Did you train your reward model on your own or use an off-the-shelf approach? Justify your decision.
	\item If your reward models are outcome-based, justify why focusing only on the terminal state is sufficient for your task.
    \begin{enumerate}
    	\item Are the terminal states of your reasoning trajectories always objective and determinate? If not, how do you justify outcome-based reward modeling?
    \end{enumerate}
	\item If your reward models are process-based, justify why considering intermediate reasoning steps is necessary for your task.
    \item How are the rewards designed---numerical, heuristic, generative, etc.? Justify why your design is appropriate for your task. 
\end{enumerate}

\end{mybox}

\subsection{Reasoning Evaluation}
Our survey shows that evaluations of LLM reasoning in HCI are primarily grounded in the subjective utility of target users (section \ref{sec:hci-eval-subj}).
Further, we also observed that many prior works rely on previously established benchmark results to motivate LLM applications.
Our reflection prompts in this section consider both these pre- and post-application approaches to evaluation to encourage HCI practitioners to reflect on how evaluation permeates in different forms.
While we also discussed that reasoning evaluation also happens during the training phase (section \ref{sec:hci-eval-abs}), we included most of these relevant prompts in the section on reasoning execution.

\begin{mybox}[Reasoning Evaluation]
\begin{enumerate}
\item Do you conduct a preliminary evaluation of LLMs' reasoning abilities prior to applying them for your task? If not, explain why this is not necessary. 
\item Do you refer to LLMs' benchmark performance to interpret their reasoning ability? If so, justify why these established results are sufficient for your task. 
\begin{enumerate}
    \item How aligned are the modes of reasoning in these benchmarks with respect to your task? Justify the level of alignment necessary for your task.
    \item Do your benchmarks contain intermediate reasoning steps? Justify why this does (not) affect your interpretation of LLM reasoning required for your task.
	\item How do you justify that the LLMs you use did not memorize the benchmark test datasets? 
\end{enumerate}
\item If your evaluation of LLM reasoning is predominantly qualitative, justify why this is necessary and/or sufficient?
\item Do you evaluate LLMs' reasoning abilities in relation to how they benefit your target users? If so, justify why LLM reasoning is necessary for your target user.
\begin{enumerate}
	\item How do you ensure and assess whether your users consider LLMs' reasoning behavior when determining your system's utility? Justify your method.
	\item How do you evaluate whether users consider the reasoning process of LLMs as a whole or in parts (such as reasoning steps)? Justify your method.
	\item Do users consider only the terminal state of reasoning or the intermediate steps too? Justify your assessment.
\end{enumerate}
\item If your evaluation involves quantitative methods, are they against ground truth values or user feedback? Justify why you chose one over the other, if applicable.
\end{enumerate}
\end{mybox}

\section{Limitations and Conclusions}
The use of LLMs in HCI has increased exponentially in recent years and can only be expected to increase (at least) in the near future. 
In contrast to other technologies, LLMs are getting integrated into HCI research in unique ways due to their ability to generate human-like responses.
One such unique characteristic is their presumed reasoning behavior that shapes how HCI practitioners imagine their capabilities, interpret their outcomes, and use them in diverse contexts.
Despite this central role of LLM reasoning, this work argued how HCI decontextualizes it from its foundational structural and methodological characterizations, leading to abstractions of both its capabilities and limitations.
To empirically support our hypothesis, we relied on a literature survey of 258 LLM-related papers published at CHI in the last six years.

However, our analysis is not without limitations. While we used an exact keyword matching on title and abstract to create our corpus, we may have missed relevant works that engage significantly with LLMs without mentioning them in these two fields.
Further, our keywords were based on a prior work in 2024, and due to the pace at which LLM research is growing, papers published in 2025 may have employed different terminology, leading to further omissions.
We also relied on ACM Digital Library's search functionality and did not test its robustness by cross-checking papers it did not return as LLM-relevant.
Due to all these factors, our corpus may have high precision but lower-than-ideal recall. 
Importantly, we deliberately chose not to use LLMs to code, since they are the very object of our investigation.
This required several days of careful manual coding by the first two authors, though, as with all manual annotation, the possibility of mislabeling remains.

That being said, we believe our corpus is sufficiently representative to capture how LLM reasoning is approached in HCI and to reveal the overall patterns identified in this study.
Beyond surveying HCI research, we also condensed and organized the contributions from NLP/ML on LLM reasoning and related them coherently to the three parts of our research question---how reasoning is framed, executed, and evaluated---and elaborated on the relevant elements within each that remain obscured in HCI discourse. 
As a constructive step forward, we developed an extensive list of reflection prompts, drawing on our survey and synthesis of NLP/ML literature, to support practitioners in critically engaging with LLM reasoning before applying it in their work.
In line with similar efforts, we believe our research contributes to bringing the NLP/ML and HCI communities together towards an informed and reflective use of LLM reasoning.



\bibliographystyle{ACM-Reference-Format}
\bibliography{references}

\appendix
\newpage
\section{Full Query Syntax}
We collected all CHI papers published between 2020-2025 from the ACM Digital Library using the following query:

\begin{lstlisting}
"query": { Title:("language model" OR "llm" OR "foundation model" OR "foundational model" OR "GPT" OR "ChatGPT" OR "Claude" OR "Gemini" OR "Falcon") OR Abstract:("language model" OR "llm" OR "foundation model" OR "foundational model" OR "GPT" OR "ChatGPT" OR "Claude" OR "Gemini" OR "Falcon") }
"filter": { Conference Collections: CHI: Conference on Human Factors in Computing Systems, Article Type: Research Article, E-Publication Date: (01/01/2020 TO 12/31/2025) }
\end{lstlisting}

\section{Anonymized Project Repository}

We open-source all our annotation data here: \url{https://anonymous.4open.science/r/llm-reasoning-chi-86CA}. Further, the repository also contains the structured list of reflection prompts proposed in our discussion.

\section{Final Codebook}

Table \ref{tab:code} shows our final codes and their descriptions.

\begin{table*}
    \centering
    \resizebox{\textwidth}{!}{%
    \begin{tabular}{|l|l|l|}
    \hline
     \textbf{Category} & \textbf{Codes}  & \textbf{Description} \\ \hline
      Year (6 codes)  & \begin{tabular}[c]{@{}l@{}} \\ [-2mm] 2020, 2021, 2022, 2023, 2024, 2025 \\ [2mm]\end{tabular} & The year of study \\
      
      Domain (11 codes)  & \begin{tabular}[c]{@{}l@{}} \\ [-2mm] Communication \& Writing, Augmenting Capabilities, Education, \\ Responsible Computing, Programming, Reliability \& Validity of LLMs, \\ Well-being \& Health, Design, Accessibility \& Aging, Creativity, Others \\ [2mm]\end{tabular} &  The application domain \\ 
      
      Roles (5 codes)  & \begin{tabular}[c]{@{}l@{}} \\ [-2mm] LLMs as system engines, LLMs as research tools, \\ LLMs as participants
        \& users, LLMs as objects of study, \\ Users' perceptions of LLMs \\ [2mm]\end{tabular} & Roles of LLMs \\
        
      Reasoning (2 codes)  & \begin{tabular}[c]{@{}l@{}} \\ [-2mm] Reasoning word, Reasoning mention \\ [2mm]\end{tabular} & \begin{tabular}[c]{@{}l@{}} \\ [-2mm] The presence of the word \\ ``reasoning'', Mention of \\ LLMs' reasoning abilities\\ [2mm]\end{tabular} \\

     LLM Task (25 codes) & \begin{tabular}[c]{@{}l@{}} \\ [-2mm] 
     analyzing bias / quality / errors, assisting coding / debugging, \\ assisting writing, explaining concepts / clarifying, extracting / \\ retrieving information, generating code / pseudocode, generating \\
     creative content (poems, analogies, metaphors, fiction), generating \\ dialogue / conversation, generating ideas, generating multimodal \\ 
     content (images, video, 3d, design), generating q/a, generating \\
     stories / narratives, generating text, guiding user actions / \\ 
     behavior, optimizing prompts / responses, planning tasks / \\
     strategies, providing educational support (teachers / students / \\ pedagogy), providing medical support / advice, providing mental \\
     health support, providing recommendations, recognizing objects / \\ classifying content, simulating personas / behaviors, summarizing \\ 
     text, supporting collaboration / communication, transforming / \\
     rewriting text
     \\ [2mm]\end{tabular} & The primary task of LLMs \\ 

     LLM Name (40 codes) & \begin{tabular}[c]{@{}l@{}} \\ [-2mm] 
     AutoGen, CareCall, Doubao-pro-128k, FLAN-T5, FLAN-T5-Base, GutGPT, \\
     alexa, bard, chatgpt, claude v1.3, claude-3.5-sonnet, cnn, codex, \\
     copilot, dall-e-2, dall-e-3, gemini, glm-4, gpt-2, gpt-3, gpt-3.5, \\
    gpt-3.5-turbo, gpt-4, gpt-4-turbo, gpt-4o, gpt-4o-mini, instruct-gpt,\\
    langchain, llama-3-chinese, llama-3.1-70b, llama-3.1-8b, lstm, meena,\\
    midjourney, mistral-7b, mixtral-7b, palm, qwen-2, sbert, stable diffusion \\ [2mm]\end{tabular} & LLM name \\

    Use Study (9 codes) & \begin{tabular}[c]{@{}l@{}} \\ [-2mm] 
     between-subject, diary study, expert review, focus group, interview, \\
     survey, think-aloud, within-subject, workshop \\ [2mm]\end{tabular} & User study method \\

    Stage (4 codes) & SFT, Inference, RL, Decoding & \begin{tabular}[c]{@{}l@{}} \\ [-2mm] Method of eliciting\\
     reasoning from LLMs \\ [2mm]\end{tabular} \\

    Reasoning Proxy (10 codes) & \begin{tabular}[c]{@{}l@{}} \\ [-2mm] 
    creative, explain, interpret, justify, plan, rationalize, \\
   reason, reflect, think, understand \\ [2mm]\end{tabular} & Reasoning-related concepts \\

   Reasoning Usage (6 codes) & \begin{tabular}[c]{@{}l@{}} \\ [-2mm] 
  R-motivation, R-outcome, R-critical, R-minor, R-step, R-multi \\ [2mm]\end{tabular} & 
   \begin{tabular}[c]{@{}l@{}} \\ [-2mm] 
    Reasoning used as motivator, \\ reasoning in outcome, \\ reasoning task is critical \\ reasoning task is minor \\ reasoning involves steps \\ reasoning from systems \\ [2mm]\end{tabular} \\
    
      Misc. (2 code) & Applies LLM, LLMs for Eval & \begin{tabular}[c]{@{}l@{}} \\ [-2mm] The study applies LLMs \\ to build systems, \\ LLMs are used in \\ evaluation\\ [2mm] \end{tabular}
      \\ \hline 
    \end{tabular}
    }
    \caption{Final codes of our annotation process and their descriptions.}
    \label{tab:code}
\end{table*}

\end{document}
\endinput